\documentclass[a4paper,fleqn]{cas-dc}
\usepackage[justification=centering]{caption}
\usepackage{graphicx} 
\usepackage{cas-common}
\usepackage[numbers]{natbib}

\usepackage{algorithm}
\usepackage{algorithmicx}
\usepackage{algpseudocode}
\usepackage{amsmath}
\begin{document}
\captionsetup[figure]{labelfont={bf},name={Fig.},labelsep=period}
\let\WriteBookmarks\relax
\def\floatpagepagefraction{1}
\def\textpagefraction{.001}
\shorttitle{Path-Based Reasoning over Heterogeneous Networks for Recommendation}
\shortauthors{Junwei Zhang et~al.}

\title [mode = title]{Path-Based Reasoning over Heterogeneous Networks for Recommendation via Bidirectional Modeling}



\author[1,2]{Junwei Zhang}
\address[1]{Key Laboratory of Dependable Service Computing in Cyber Physical Society (Chongqing University), Ministry of Education, Chongqing, 401331, China}
\address[2]{School of Big Data and Software Engineering, Chongqing University, Chongqing, 401331, China}

\author[1,2]{Min Gao}
\cormark[1]
\ead{gaomin@cqu.edu.cn}

\author[3]{ Junliang Yu}
\address[3]{School of Information Technology and Electrical Engineering, The University of Queensland, Queensland, 4072, Australia}

\author[4]{ Linda Yang}
\address[4]{School of Engineering, University of Portsmouth, Portsmouth, PO1 3AH, UK}

\author[1,2]{ Zongwei Wang,}


\author[1,2]{ Qingyu Xiong}

\cortext[cor1]{Corresponding author}


\begin{abstract}
Heterogeneous Information Network (HIN) is a natural and general representation of data in recommender systems. Combining HIN and recommender systems can not only help model user behaviors but also make the recommendation results explainable by aligning the users/items with various types of entities in the network. Over the past few years, path-based reasoning models have shown great capacity in HIN-based recommendation. The basic idea of these models is to explore HIN with predefined path schemes. Despite their effectiveness, these models are often confronted with the following limitations: (1) Most prior path-based reasoning models only consider the influence of the predecessors on the subsequent nodes when modeling the sequences, and ignore the reciprocity between the nodes in a path; (2) The weights of nodes in the same path instance are usually assumed to be constant, whereas varied weights of nodes can bring more flexibility and lead to expressive modeling; (3) User-item interactions are noisy, but they are often indiscriminately exploited. To overcome the aforementioned issues, in this paper, we propose a novel path-based reasoning approach for recommendation over HIN. Concretely, we use a bidirectional LSTM to enable the two-way modeling of paths and capture the reciprocity between nodes. Then an attention mechanism is employed to learn the dynamical influence of nodes in different contexts. Finally, the adversarial regularization terms are imposed on the loss function of the model to mitigate the effects of noise and enhance HIN-based recommendation. Extensive experiments conducted on three public datasets show that our model outperforms the state-of-the-art baselines. The case study further demonstrates the feasibility of our model on the explainable recommendation task.
\end{abstract}




\maketitle

\section{Introduction}
Networks such as social networks, biological networks, and transportation networks connect all kinds of data in our lives. The objects and interactions in the real world are often multi-modal and multi-type. To capture and utilize such heterogeneity of nodes and links, some researchers propose heterogeneous information networks (HIN) in many practical network mining scenarios, especially in recommender systems (RS). HIN-based models can not only alleviate data sparsity problem and improve the performances of models but also make the recommendation results explainable \cite{DBLP:journals/corr/abs-1804-11192, DBLP:conf/iui/KoukiSPOG19, DBLP:conf/chi/KunkelDMB019, DBLP:conf/kdd/Ribeiro0G16}, because the connections in HIN can intuitively reflect the extra connectivity information between users and items. Besides, these connections provide algorithm designers with a new method of debugging and improve the transparency of the recommendation model and users' cohesion. Due to their significant advantages, HIN-based RS have received extensive attention from academia and industry.

To explore the potential of HIN in RS, one line of research pays attention to making recommendations using embedding models \cite{DBLP:journals/tkde/ShiHZY19, DBLP:conf/kdd/CenZZYZ019, DBLP:conf/aaai/LinLSLZ15, DBLP:conf/cikm/Yu0LYL18}, which can be divided into two categories: node similarity and path similarity based models. The basic idea of the former is to align heterogeneous graphs in a regularized vector space and reveal the similarity between nodes by calculating the distance between representations of nodes, such as TransE \cite{DBLP:conf/aaai/LinLSLZ15} and node2vec \cite{DBLP:conf/kdd/GroverL16}. Although these models have achieved some performance improvement, they lack the consideration of discovering multi-hop relational paths. Another embedding research in RS mainly integrates heterogeneous nodes and edges with predefined path schemes and calculates the similarity between paths to learn the low-dimensional representation vector of nodes, such as meta2vec \cite{DBLP:conf/kdd/DongCS17} and IF-BPR\cite{DBLP:conf/cikm/Yu0LYL18}. Despite their prevalence and effectiveness, they can hardly cover all real-world situations because the design of the path schemes often requires the extensive domain knowledge. Briefly, the above-mentioned models based on network embedding restrict the power of node relationships in HIN.

Some researchers are aware of the adverse factors of HIN embedding and propose to conduct explicit reasoning over HIN to make recommendations instead of merely embedding the network as vectors for similarity matching. Reasoning is a process of deducing new knowledge from prior knowledge. Unlike the models based on HIN embedding, the reasoning model over HIN usually use the random walk to construct node sequences from HIN, select some nodes in the path as prior knowledge, and predict the unknown nodes drawing on sequence modeling. With the rise of machine learning, some studies have made attempts to introduce machine learning technology into the field of reasoning, such as using Markov chain \cite{DBLP:conf/recsys/HeKM17, DBLP:conf/icdm/HeM16, DBLP:conf/www/RendleFS10}, recurrent neural networks \cite{DBLP:journals/corr/HidasiK17, DBLP:conf/cikm/HidasiK18, DBLP:conf/recsys/TanXL16}, and attention mechanisms \cite{DBLP:conf/nips/VaswaniSPUJGKP17, DBLP:conf/kdd/LiuZMZ18, DBLP:journals/corr/abs-1808-06414, DBLP:conf/cikm/LiRCRLM17}. Although the above methods have made some improvements in the recommendation performance, they still have a significant limitation. Concretely, most of these models use unidirectional models from left to right to learn the node sequences of HIN, resulting in the inability to simulate complex relationships between nodes. The complex relationships in HIN can be attributed to: (1) The front and back nodes in the path. (2) The importance of nodes in a path is likely to be different. (3) User-item interactions are noisy because users may misclick some items when they are browsing items. Therefore, how to highlight these issues in the reasoning process is crucial.

In this paper, we propose an \textbf{a}ttention-based \textbf{b}idirectional \textbf{L}ong-Short Term Memory Network (LSTM) with \textbf{a}dversarial learning for recommendation model based on path reasoning over \textbf{H}IN, named \textbf{ABLAH}. Specifically, we utilize the connection information between users and items as well as other auxiliary information, such as the singer, the album, and the user’s friends to build a HIN, and set the user as the initial node to extract paths by using random walks. Then we take the acquired path as input and use the bidirectional LSTM to model path. Furthermore, considering the different importance of each node in the path can improve the performance of the model, which has been confirmed in text sequence modeling \cite{DBLP:conf/naacl/DevlinCLT19}. We adopt the attention mechanism to focus on the importance of different nodes. Compared with unidirectional reasoning, the bidirectional model is more suitable for modeling the complex relationships of the nodes. However, due to noisy relationships in HIN, it is not straightforward to train the bidirectional model for reasoning. In view of this, we develop an optimized adversarial model, where adversarial regularization terms are added to the loss function. With the attention-based bidirectional LSTM as the main recommender model, a.k.a ABLH, we design a minimax function, which extends the loss of the ABLH by introducing adversarial regularization terms, which is learned by minimizing the original loss function and added to the embedding representation learned from the ABLH. In this way, we can alleviate the problem of high time complexity caused by repeated training of the ABLH model. Extensive experiments on three real-world data sets demonstrate the effectiveness of our method. In summary, we have made the following contributions:

\begin{enumerate}[\textbullet]
\item We design a novel path-based reasoning approach using attention-based bidirectional LSTM for recommendation over HIN. 
\item To the best of our knowledge, we are the first to combine the adversarial regularization term and the embedding of HIN to alleviate the problem of noise relationships in recommendation.
\item We conduct extensive experiments on multiple public 
datasets to demonstrate the superiority of our model. The results show that it outperforms some state-of-the-art recommendation models, and the case study shows the feasibility of our model on the explainable recommendation task.
\end{enumerate}

In the rest of this paper, we further discuss the related work in Section 2. In Section 3, we present our ABLAH in detail. We describe experimental research in Section 4. Finally, we summarize our work and look forward to the future work in Section 5.

\section{Related Work}
\subsection{Recommendation Models Based on HIN}
As an emerging direction, HIN contains abundant information about users and items in RS, where nodes and edges are different types. Some researchers realize the importance of HIN in recommendation models. Wang et al. \cite{DBLP:conf/kdd/FengW12} propose to use multiple types of information to build a HIN, mapping different types of edges to different feature sets, and learning the weight of each edge. Some authors \cite{DBLP:conf/recsys/YuanCZ11} use social information to explore the social relationship of the membership, and use two fusion strategies, regularization and collective matrix decomposition, in HIN-based recommendation models. Besides, Sun et al. \cite{DBLP:journals/pvldb/SunHYYW11} introduce the concept of meta-path similarity to a HIN-based model. Yu et al. \cite{DBLP:conf/cikm/Yu0LYL18} identify more reliable friends with similar preferences by designing meaningful meta-paths for users to optimize the social BPR model \cite{DBLP:conf/cikm/ZhaoMK14}. Similarly, some studies \cite{DBLP:conf/wsdm/ChenS17, DBLP:conf/kdd/DongCS17, DBLP:journals/corr/HuangM17, DBLP:journals/tkde/ShiHZY19} also use the random walk technique based on meta-path to construct the sequences of nodes, and learn the vector representation of nodes to improve the performance of the RS. In recent years, with the rise of graph neural networks, some researchers take HIN graph data as an input, design graph neural network architectures to generate embedding representations for each node, and consider the impact of different adjacent groups for RS \cite{DBLP:conf/kdd/ZhangSHSC19, DBLP:conf/www/WangJSWYCY19, DBLP:conf/kdd/FanZHSHML19}. Most of these methods rely on path-based similarity, whereas the paths should be designed in advance. It is not practical to cover all possible meta-paths in large-scale HIN.

\subsection{Recommendation Models Based on Path Reasoning}
The graph-based RS contain various types of data 
sources, which can enrich the interactive information. The methods of reasoning implicit relationship among nodes in the graph can not only improve the accuracy of the RS but also further explain the recommendation results \cite{DBLP:conf/cikm/HeCKC15, DBLP:conf/icde/HeckelVPD17, DBLP:conf/www/Wang0FNC18}. There are several typical recommendation models based on the path reasoning. For example, Wang et al. \cite{DBLP:conf/aaai/WangWX00C19} propose to use LSTM to simulate the sequence of nodes in paths, which are selected from the knowledge graph. This model captures the order dependence of nodes to infer users’ preferences. The authors conduct experiments on multiple datasets to verify the effectiveness of the proposed model. Moreover, unlike existing methods that only focus on using knowledge graphs for more accurate recommendations, Xian et al. \cite{DBLP:conf/sigir/XianFMMZ19} propose a policy guided path reasoning method (PGPR) that uses reinforcement learning to reason on knowledge graphs and explain recommendation results. It is a flexible graph reasoning framework that can be extended to many other graph-based tasks. With the help of reasoning, multi-source heterogeneous data can be integrated into the RS, and significant performance improvement has been achieved. However, these models based on path reasoning are still at an initial stage of development and face many challenges.

\subsection{Recommendation Models Based on Adversarial Learning}
Recently, generative adversarial networks (GANs) have led a revolution in the field of deep learning. They adopt adversarial learning, in which are added adversarial disturbances to the input data or model parameters to improve the robustness of RS. The first model that applies GANs to RS is IRGAN \cite{DBLP:conf/sigir/WangYZGXWZZ17}, which attempts to unify the generative and discriminative information retrieval models under a framework. He et al. \cite{DBLP:conf/sigir/0001HDC18} verify that adding adversarial disturbances to RS can improve the robustness of the original system, and the authors propose APR, which urges the model to consider the deviation caused by noise in advance. Tang et al. \cite{DBLP:journals/corr/abs-1809-07062} extend APR into the field of image recommendation to improve the robustness of image recommendation. Furthermore, some researchers have tried to combine adversarial learning with RS in other fields, such as music and movies \cite{DBLP:conf/icde/TongLZSC19, DBLP:conf/sigir/TranSL19, DBLP:conf/ijcnn/YuanYB19}. 

\section{The Proposed Model}
In this section, we introduce the structure of our model. First, we formalize the notations of HIN and the HIN-based recommendation task. Then we introduce the various components of the proposed model, ABLAH, which consists of ABLH and adversarial regularization term. Finally, we discuss the model size and the time complexity of our model. Fig.~\ref{f2} shows the schematic overview of ABLAH. 
\subsection{Preliminary}
As a special kind of network, HIN includes multiple types of nodes and edges. In this subsection, we introduce the main notations involved in this paper and formalize the problem of HIN-based recommendation.

\begin{figure}[pos=htp]
\centerline{\includegraphics[scale=0.15]{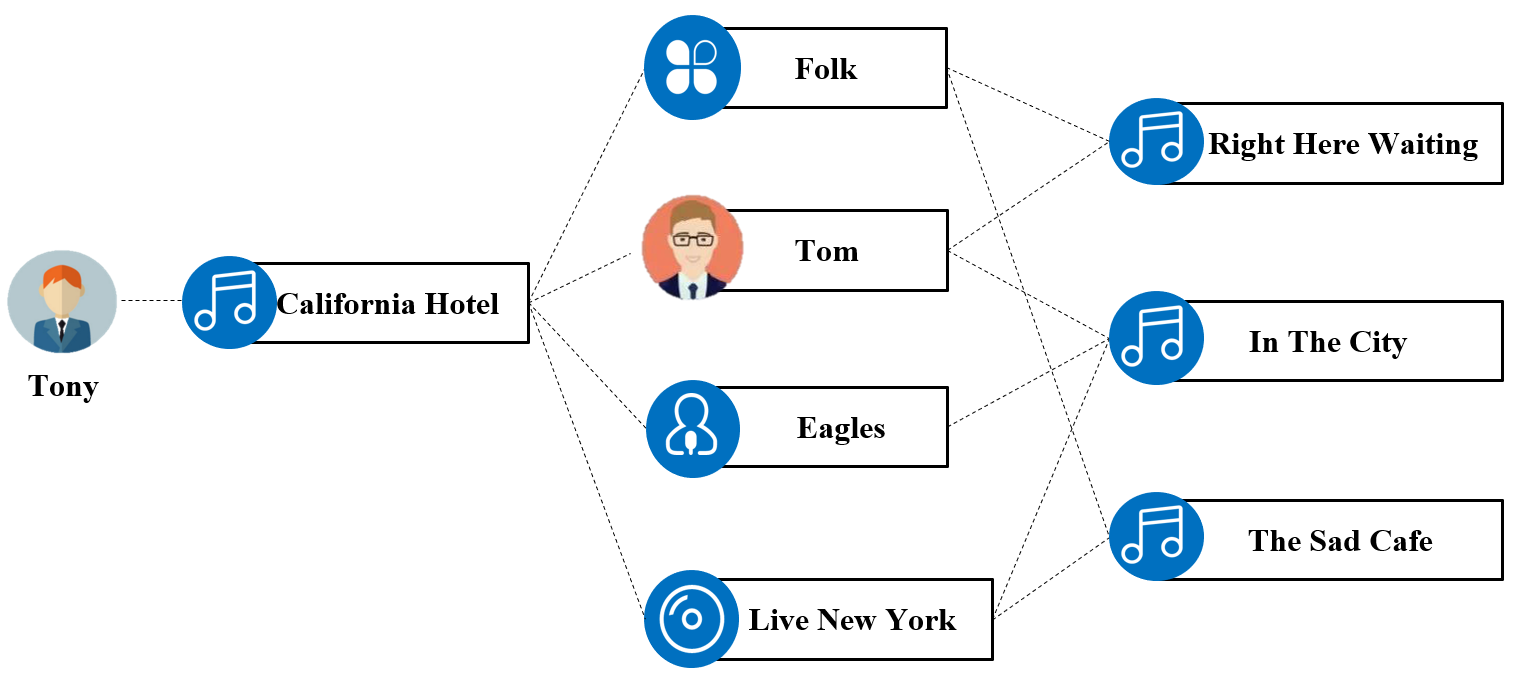}}
\caption{Illustration of recommendation in the music domain. The dashed lines between nodes are the corresponding relations.}
\label{f1}
\end{figure}

\textbf{Heterogeneous Information Networks (HIN).} We formally use $G=(V, E)$ to represent heterogeneous information networks (HIN), where $V$ is the set of nodes and $E$ is the set of edges. In the HIN, each node $v$ and each edge $e$ have a mapping relationship $\varphi(v):V\rightarrow T_{V}$, $\varphi(e):E \rightarrow T_{E}$, $T_{V}+T_{E}\geq 3$, where $T_{V}$ and $T_{E}$ are the set of types of nodes and edges, respectively.

\textbf{Path in HIN.} Within $G$, we treat the node sequences from user $u$ to item $i$ as a path, which is defined as $p=[v_{1}, v_{2},\cdots ,v_{L}]$, where $L$ is the maximum number of nodes in a path. Different from the handcrafted meta-path, we fix the user node $u$ and the target node $i$ as the first and the last node in the path respectively, whereas other nodes are extracted by random walk. An example of HIN about music is illustrated in Fig.~\ref{f1}. There are three paths from the same user $Tony$ to the same song $In The City$, which can clearly express their different multi-step relationships and reveal different reasons why the RS recommends this song to the user.

\begin{enumerate}[\textbullet]
\item $p_{1}=[\emph{Tony}, \emph{California Hotel}, \emph{Tom}, \emph{In The City}]$, 
\item $p_{2}=[\emph{Tony}, \emph{California Hotel}, \emph{Eagles}, \emph{In The City}]$, 
\item $p_{3}=[\emph{Tony}, \emph{California Hotel}, \emph{Live New York}, \emph{In The City}]$, 
\end{enumerate}

\textbf{HIN-Based Recommendation Task.} The path set can be formalized as $P(u,i)=\left \{ p_{1}, p_{2},\cdots, p_{k} \right \}$, which take the target user $u$ as the head node and the target item $i$ as the tail node. The task is to estimate the probability that the user will buy or like the item. We obtain the probability through the paths between the user and the item, which can be computed as
\begin{equation}
\label{eq:eq1}
\begin{aligned}
\hat{y}_{ui}=f_{\theta }\left ( u,i | P\left ( u,i \right ) \right ),
\end{aligned}
\end{equation}
where $\hat{y}_{ui}$ represents the interaction probability between the user and the item, and $f$ denotes the mapping function with the parameter $\theta$. 

\subsection{The Architecture of ABLH}
\begin{figure*}[pos=htp]
\centerline{\includegraphics[width=6.2in, height=2.5in]{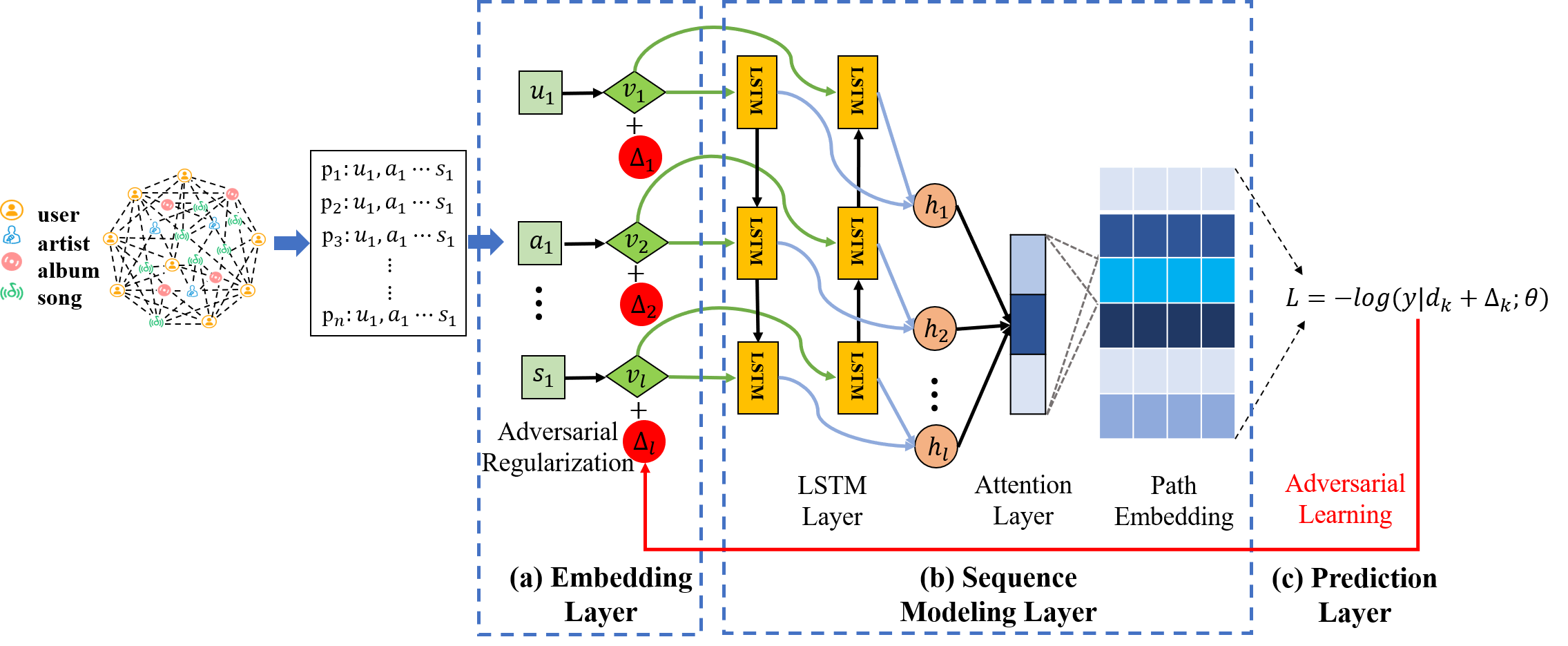}}
\caption{The overview of our proposed framework for enhanced HIN-based recommendation. (a) Embedding layer. (b) Sequence modeling layer. (c) Prediction layer.}
\label{f2}
\end{figure*}

The main idea of this paper is to model node sequence information of HIN and capture the complex relationships between nodes in HIN. Hence, we firstly focus on the attention-based bidirectional LSTM for reasoning and introduce the basic solution (ABLH) to achieve this goal. The input of ABLH is a paths set about each user-item pair, and the output is a score of probability that the user might interact with the target item. As illustrated in the Fig.~\ref{f2}, our model contains three main parts: (a) \textbf{The Embedding Layer:} This module maps the value and type of the node into low-dimensional vector representations to obtain the initialization vectors for the downstream task; (b) \textbf{The Sequence Modeling Layer:} We adopt the attention-based bidirectional LSTM to capture the contextual relationship between the nodes in a path. To be specific, both the predecessor node and the subsequent nodes on the path may influence each other and the weights of nodes in the same path could be different; (c) \textbf{The Prediction Layer:} To combine the representation of nodes from two-way sequences and predict the probability of user interaction on the target item, we adopt two-layer feed-forward neural network to obtain the score.

\textbf{Embedding Layer.}
We use random walk to extract $K$ paths , which contain different types of nodes and edges. In each path, the head is the user $u$ and the tail is the item $i$. We map types of nodes and the specific value of nodes into two vectors $e_{l}^{t} \in \mathbb{R}$ and $e_{l}^{s} \in \mathbb{R}$. Each representation captures different meanings with respect to different perspectives. To fully exploit the nodes' type and specific value, we concatenate these two embedding to form the final node embedding $h_{l}$. For each node in the path, the new initialization vector of the node $h_{l}$ can be obtained according to Eq.~\ref{eq:eq2}:
\begin{equation}
\label{eq:eq2}
\begin{aligned}
h_{l}=e_{l}^{t} + e_{l}^{s}.
\end{aligned}
\end{equation}
Therefore, through the embedding layer, we can obtain an embedding $[h_{1}, h_{2},\cdots ,h_{L}]$ for the path $p_{k}$, where each element represents the vector of a node.

\textbf{Sequence Modeling Layer.}
Through HIN-based embedding, we obtain the vector of nodes. For each user-item pair, we take the embedding of nodes $h_{l}$ in paths as input, and the probability of user's favorite item is obtained through the attention-based bidirectional LSTM. We use the attention-based bidirectional LSTM to further infer the final target item due to its capability of capturing the complex sequence information deeply among nodes. Different from the sentences in the NLP problem, the number of nodes in path of HIN is finite. We need to pay more attention to the importance of different nodes in the path and their impact on the entire path. Compared with traditional unidirectional modeling model, our bidirectional one considers both left and right context in a path to achieve context-based inference. As shown in Fig.~\ref{f2}, the output of the model is jointly determined by the hidden state of the forward LSTM and backward LSTM. Specifically, in the forward LSTM, the target item $i$, which is the last node in the path $p$, is represented as $h_{L}$. The low-dimensional vector of the previous node is represented as $h_{L-1}$. The hidden state of the node is $\overrightarrow{h}_{L-1}$ and its cell state vector is defined as $\overrightarrow{c_{L-1}}$. We define Eq.~\ref{eq:eq3} to learn the hidden state of target item $\overrightarrow{h}_{L}$: 

\begin{equation} \nonumber
\begin{aligned}
\overrightarrow{z_{L}}=tanh\left ( \overrightarrow{W_{z}}h_{L-1}+ \overrightarrow{W_{h}}\overrightarrow{h}_{L-1}+\overrightarrow{b_{z}} \right )
\end{aligned}
\end{equation}

\begin{equation}
\label{eq:eq3}
\begin{aligned}
& \overrightarrow{f_{L}}=\sigma \left ( \overrightarrow{W_{f}}h_{L-1}+ \overrightarrow{W_{h}}\overrightarrow{h}_{L-1}+\overrightarrow{b_{f}} \right ) \\ 
& \overrightarrow{i_{L}}=\sigma \left ( \overrightarrow{W_{i}}h_{L-1}+\overrightarrow{W_{h}}\overrightarrow{h}_{L-1}+\overrightarrow{b_{i}} \right ) \\ 
& \overrightarrow{o_{L}}=\sigma \left ( \overrightarrow{W_{o}}h_{L-1}+\overrightarrow{W_{h}}\overrightarrow{h}_{L-1}+\overrightarrow{b_{o}} \right )  \\
& \overrightarrow{c_{L}}=\overrightarrow{f_{L}}\ast \overrightarrow{c_{L-1}}+\overrightarrow{i_{L}}\ast \overrightarrow{z_{L}} \\
& \overrightarrow{h}_{L}=\overrightarrow{o_{L}}\ast tanh(\overrightarrow{c_{L}}),
\end{aligned}
\end{equation}
where $z\in \mathbb{R}$ represents the transformation module, $\overrightarrow{i_{L}}$, $\overrightarrow{o_{L}}$, and $\overrightarrow{f_{L}}$ represent input, output, and forget gate respectively. $\overrightarrow{W_{z}}$, $\overrightarrow{W_{i}}$, $\overrightarrow{W_{f}}$, and $\overrightarrow{W_{o}}$ are mapping coefficient matrices, while $\overrightarrow{b_{z}}$, $\overrightarrow{b_{i}}$, $\overrightarrow{b_{f}}$, and $\overrightarrow{b_{o}}$ are bias vectors. $\sigma (\cdot )$ is the activation function, and $(\ast)$ means the elements-wise multiplication. The backward LSTM layer only needs to take the opposite node sequences as input to obtain the hidden state $\overleftarrow{h}_{L}$. In this way, we can make full use of the forward and backward information in the path. Finally, the representation vector of $h_{L}$ is calculated by concatenating the hidden state vectors generated in two directions, as shown in Eq.~\ref{eq:eq4}:

\begin{equation}
\label{eq:eq4}
\begin{aligned}
\overline{h}_{L}=\left [ \overrightarrow{h}_{L}\oplus \overleftarrow{h}_{L} \right ].
\end{aligned}
\end{equation}

Obviously, not all nodes contribute equally to the representation of a path, some nodes are more important than others. Yuan et al. \cite{DBLP:journals/isci/YuanWYLL20} have set a good example to capture the sequential information among different nodes in RS by using LSTM, but they did not consider the importance of nodes in a path. Hence, we highlight those valuable nodes via attention mechanism and the enhanced path representation is computed as shown in Eq.~\ref{eq:eq4}:
\begin{equation}
\label{eq:eq5}
\begin{aligned}
& M_{u}=tanh\left ( \overline{H_{u}} \right ),  \\ 
& \alpha_{u} = softmax\left ( W_{u}M_{u} \right ),  \\
& R_{u}=\overline{H}_{u}\alpha_{u}^{T},  \\  
\end{aligned}
\end{equation}
where $\overline{H_{u}}$ is the hidden state representation matrix of all path nodes for the user $u$, $\alpha_{u}$ is the attention matrix, and $W_{u}$ is the coefficient matrix. We use weighted sum of the output vectors to represent the vector of the path $R_{u}$.

We first use a two-layer fully connected network to further optimize the representations, as shown in Eq.~\ref{eq:eq6}:
\begin{equation}
\label{eq:eq6}
\begin{aligned}
s_{k}=W_{1}^{T}ReLu\left ( W_{2}^{T} R_{u} \right ),         
\end{aligned}
\end{equation}
where $W_{1}^{T}$ and $W_{2}^{T}$ are the coefficient matrix of the forward neural network respectively. Since there are multiple paths between the user $u$ and the item $i$, we average the path vectors as the final representation vector of the path $s_{ui}$:

\begin{equation}
\label{eq:eq7}
\begin{aligned}
s_{ui}=\frac{1}{K}\sum_{k=1}^{K}s_{k}.         
\end{aligned}
\end{equation}

\textbf{Prediction Layer.}
By stacking bidirectional sequence modeling layer, the path representation are capable of receiving the information propagated from right and left node sequences. $\hat{y}_{ui}$ is probability score of user interaction on items in each path. Given an instance $(u,i)$, $\hat{y}_{ui}$ can be computed as follows:
\begin{equation}
\label{eq:eq8}
\begin{aligned}
\hat{y}_{ui}=\sigma \left ( s_{ui} \right ),
\end{aligned}
\end{equation}
where $\sigma$ is the activation function. 

Similar to the work of He et al. \cite{DBLP:conf/sigir/0001HDC18}, we consider the recommendation task as a classification problem in which the target item is the label. Therefore, we design the following loss function to learn the parameters of our model:
\begin{equation}
\label{eq:eq9}
\begin{aligned}
L=-logp \left ( y|s_{ui};\theta \right ),
\end{aligned}
\end{equation}
where $\theta$ denotes the parameters of the sequence modeling model.

\subsection{Enhance Framework with Adversarial Learning}
We note that it is not straightforward to train the ABLH for reasoning. This is mainly because the above model ignores the effect of noise in the interaction data. In RS, users sometimes click on items beyond their interests, which leads to noisy edges that should not exist in HIN. To cope with the noise problem and improve the accuracy of the node representations, we propose ABLAH, which refines ABLH with adversarial learning. Concretely, inspired by the existing models based on adversarial learning \cite{DBLP:conf/sigir/0001HDC18, DBLP:journals/corr/abs-1809-07062}, we design a new loss function and optimize it to achieve the above purpose. Typically, the adversarial regularization term is applied to either feature representation or model parameters. An intuitive choice is to apply term to model parameters. However, this method is difficult to add the regularization term to the appropriate parameter position during the training process because the model structure is complicated. Moreover, the latter solution tend to cause the model over-fitting problem because the interaction information is sparse and vast. To avoid these difficulties, we add the adversarial regularization term to the node embedding representation vector. Explicitly, we define the objective function as Eq.~\ref{eq:eq10}:
\begin{equation}
\label{eq:eq10}
\begin{aligned}
& L_{adv}=-log p \left ( y|s_{ui}+\Delta _{k};\theta  \right ), \\
& where, \Delta _{k}=-\epsilon g/\left \| g \right \|_{2}, \\
& g=\bigtriangledown _{s_{ui}}logp\left ( y|s_{ui};\theta  \right ),
\end{aligned}
\end{equation}
where $\Delta _{k}$ represents the adversarial regularization term, $\epsilon $ controls the size of $\Delta _{k}$, and the loss function is finally calculated by back-propagation.

\begin{algorithm}[htp]
  \caption{ ABLAH}
  \label{alg:Framwork}
  \begin{algorithmic}[1]
    \Require
      network $G=(V, E)$, node embedding dimension $d$, learning rate $\lambda$, the number of paths per user $K$, the number of nodes in each path $L$, coefficient $\epsilon$, the number of neurons $r$, the batch size $b$, and the number of iterations $iter$;
    \Ensure
      Ensemble of classifiers on the current batch, $E_n$;
    \State Initialize all the model parameters $\theta$, initialize the coefficient matrix $W$ and bias vector $b$;
    \State Utilize random walks to generate paths for each user, where the head node is $u$ and the tail node is $i$;
    \State Initialize the node vector $h_{l}$;
    \Repeat
    	\State Adopt the bidirectional LSTM model to achieve path-based reasoning over HIN using Eq.~\ref{eq:eq3} and Eq.~\ref{eq:eq4}; 
    	\State Adopt the attention mechanism to capture different importance of nodes using Eq.~\ref{eq:eq5};
    	\State Calculate the hidden state of the target item $\overline{h}_{L}$;
    	\State Update the coefficient matrix $W$ and bias vector $b$ using Eq.~\ref{eq:eq9};
    	\State Calculate objective function $L_{adv}$ and update the adversarial regularization term $\Delta _{k}$ using Eq.~\ref{eq:eq10}; 
    \Until{($k<K$ and $l<L$ and $u<m$)} 
  \end{algorithmic}
\end{algorithm}

\subsection{Discussion}
Our model integrates multiple components to cope with different issues. We summarize our algorithm in Algorithm ~\ref{alg:Framwork} and discuss the model size and time complexity of our model in the following section. 

\textbf{Model Size.} For the embedding layer, the size of the vector representation is $m \times d$. Besides, the sequence modeling layer has the parameter of size $k(l+1)d$, and the size of weight matrix is $d \times d$. By adding up all the numbers, the total model size is $2md+2k(l+1)d+2 \times 6 \times d^{2}$. Considering that $d$ and $k$ are small numbers generally less than 100 and the number of layers is usually less than 3, the model size is still a small number.

\textbf{Time Complexity.} In our model, the source of time complexity mainly comes from the process of constructing HIN and random walks. For the HIN, the number of user is $m$, and the number of item is $n$. So the computation cost is $O(d^{2}k)$. The time complexity for computation through fully connected layer is $O(m^{2}k)$. The time consumption for modeling node sequences is at least $O(kld^{2})$. As can be seen, our model has extra time expenses in searching for paths of user. Considering the sparsity of the interaction matrices, the compromise is acceptable. After witnessing the improvements presented in Section 4, we insist that the small sacrifice in time complexity is worthy.

\section{Experiments}
In this section, we conduct experiments on three real-world datasets to evaluate the superiority of proposed model. We focus on answering the following research questions (RQ):
\begin{enumerate}[\textbullet]
\item \textbf{RQ1:} Compared with the traditional recommendation models and the state-of-the-art methods, how does the proposed model perform?
\item \textbf{RQ2:} How the key parts of our model affect recommendation performance and whether adversarial learning can improve the accuracy and robustness of the model?
\item \textbf{RQ3:} How do some key parameters affect the recommendation performance?
\item \textbf{RQ4:} Can the proposed model provide convincingly explanation on recommendation results?
\end{enumerate}

\subsection{Experimental Settings}
\textbf{Datasets.} We use three real-world music datasets: Nowplaying, Xiami, and Yahoo. The Nowplaying is a dataset, which is published by Twitter about users’ music listening behavior and contains 87,663 interactions with 8,820 songs. The Xiami dataset contains the listening data of 4,270 users from the Xiami Music APP. Another dataset Yahoo comes from the Yahoo Music APP, which contains some detailed descriptions of music, such as artists, albums, etc. The statistics of the datasets are shown in Table~\ref{tab:tab1}. For dataset preprocessing, we follow previous studies \cite{DBLP:conf/www/HeLZNHC17, DBLP:conf/icdm/KangM18, DBLP:conf/www/RendleFS10, DBLP:conf/wsdm/TangW18}, and we filter out the user with less than five feedback for three datasets. Besides, we group the interaction records by users and construct HIN, which is used to build node sequences. For each user, we holdout the 80\% interaction history to construct the training sets and the rest is for testing.

\begin{table}[pos=htp]
\caption{The statistics of datasets.}
\footnotesize
\label{tab:tab1}
\begin{tabular}{clllll}
\hline
\textbf{Datasets} & \multicolumn{1}{c}{\textbf{Users}} & \multicolumn{1}{c}{\textbf{Items}} & \multicolumn{1}{c}{\textbf{Artists}} & \multicolumn{1}{c}{\textbf{Albums}} & \multicolumn{1}{c}{\textbf{Density}} \\ \hline
\textbf{Nowplaying} & 155 & 8,820 & 1,704 & \multicolumn{1}{c}{/} & 6.41\% \\
\textbf{Xiami} & 4,270 & 289,083 & 32,918 & 94,969 & 0.22\% \\
\textbf{Yahoo} & 10,732 & 136,665 & 20,541 & 9,440 & 0.22\% \\ \hline
\end{tabular}
\end{table}

\textbf{Evaluation Metrics.} 
To evaluate the recommendation performance of all models, we adopt a leave-one-out evaluation mechanism, similar to the studies in \cite{DBLP:conf/www/HeLZNHC17, DBLP:conf/icdm/KangM18, DBLP:conf/wsdm/TangW18}. We employ two conventional RS evaluation metrics: Hit Ratio (HR) and Normalized Discounted Cumulative Gain (NDCG). Intuitively, the HR@K metric is a commonly used indicator to measure the recall rate while NDCG@K metric is a ranking metric. In this paper, we use $K = 5,10$ to report HR and NDCG. For both of the evaluation metrics, the higher the value, the better the performance.

\textbf{Comparison Method.}
We compare the proposed model with the following models to answer RQ1.
\begin{enumerate}[\textbullet]
\item \textbf{POP}: It is the simplest recommendation model which recommends the most popular songs to users.
\item \textbf{BPR} \cite{DBLP:conf/uai/RendleFGS09}: It adopts a pairwise ranking loss function to optimize the implicit matrix factorization model.
\item \textbf{CDAE} \cite{DBLP:conf/wsdm/WuDZE16}: It uses variational autoencoders to recommend items to users.
\item \textbf{NeuMF} \cite{DBLP:conf/www/HeLZNHC17}: It uses neural networks to model user-item interaction information.
\item \textbf{RNN4rec} \cite{DBLP:journals/corr/HidasiKBT15}: It uses the recurrent neural network to model the node sequences and recommend items.
\item \textbf{CNN4rec} \cite{DBLP:conf/wsdm/TangW18}: It models node sequences over HIN using convolutional neural network (CNN) to predict the target item.
\item \textbf{KPRN} \cite{DBLP:conf/aaai/WangWX00C19}: It uses LSTM to capture the sequential dependencies of nodes and compose the embedding of nodes and edges to construct the representations of paths.
\end{enumerate}

\textbf{Implementation Details.}
In practical applications, it is not feasible to fully explore all connection paths over HIN. As suggested by previous work \cite{DBLP:conf/icdm/KangM18}, we ignore distant connections and adopt fixed-length extraction paths that are efficient for reasoning.

For fairness, we choose the best performing parameter of all methods as the comparison parameter. For our model, we implement it through TensorFlow, where all parameters are optimally set by grid search. We use Adam \cite{DBLP:journals/corr/KingmaB14} to train the model, where the initial learning rate is 0.001, which decreases linearly with the increase of the number of training and the decrease of the loss function. For other parameters, we set the number of layers of the LSTM to 2 and the number of neurons in each layer to 128. The length of each user path is 4 in the Nowplaying dataset, 5 in the Xiami dataset, and 5 in the Yahoo dataset. We experimentally set the dimension of the low-dimensional vector as 32 and the dropout rate as 0.8. All the models are trained based on NVIDIA GeForce GTX 1080 with a batch size of 128. Our code is in https://github.com/0411tony/Yue.

\subsection{Overall Performance Comparison (RQ1)}
Table~\ref{tab:tab2} shows the best recommendation performance of all models on three datasets. In particular, the best results in the evaluation metrics are bold.
\begin{table*}[pos=htp]
\small
\caption{Performance comparison of different recommendation models.}
\label{tab:tab2}
\begin{tabular}{c|cccccccccl}
\hline
\textbf{Datasets} &
  \textbf{Metric(\%)} &
  \textbf{POP} &
  \textbf{BPR} &
  \textbf{CDAE} &
  \textbf{NeuMF} &
  \textbf{RNN4rec} &
  \textbf{CNN4rec} &
  \textbf{KPRN} &
  \textbf{ABLAH} &
  \multicolumn{1}{c}{\textbf{\begin{tabular}[c]{@{}c@{}}Improve\\ (\%)\end{tabular}}} \\ \hline
\multirow{4}{*}{\textbf{Nowplaying}} & HR@5    & 3.142 & 3.391 & 3.482 & 3.885 & 4.084  & 3.997  & 4.124  & \textbf{5.315}  & 28.83 \\
                                     & HR@10   & 2.813 & 3.064 & 3.547 & 3.961 & 4.563  & 4.076  & 4.119  & \textbf{5.143}  & 25.06 \\
                                     & NDCG@5  & 0.112 & 0.124 & 0.123 & 0.133 & 0.157  & 0.135  & 0.138  & \textbf{0.21}  & 61.54 \\
                                     & NDCG@10 & 0.094 & 0.137 & 0.135 & 0.142 & 0.164  & 0.145  & 0.139  & \textbf{0.197}  & 46.15 \\ \hline
\multirow{4}{*}{\textbf{Xiami}}      & HR@5    & 2.123 & 1.785 & 2.326 & 2.257 & 3.218  & 3.233  & 3.814  & \textbf{4.165}  & 9.19  \\
                                     & HR@10   & 1.864 & 1.505 & 1.911 & 1.874 & 2.098  & 2.413  & 2.925  & \textbf{3.478}  & 18.84 \\
                                     & NDCG@5  & 0.053 & 0.045 & 0.058 & 0.056 & 0.072  & 0.079  & 0.093  & \textbf{0.135}  & 44.44 \\
                                     & NDCG@10 & 0.045 & 0.037 & 0.047 & 0.046 & 0.057  & 0.058  & 0.062  & \textbf{0.084}  & 33.33 \\ \hline
\multirow{4}{*}{\textbf{Yahoo}}      & HR@5    & 5.085 & 8.901 & 9.125 & 9.346 & 10.237 & 11.428 & 11.884 & \textbf{12.237} & 2.95  \\
                                     & HR@10   & 4.624 & 7.421 & 7.466 & 8.158 & 8.736  & 9.473  & 10.074 & \textbf{10.875} & 7.94  \\
                                     & NDCG@5  & 0.193 & 0.311 & 0.344 & 0.349 & 0.412  & 0.509  & 0.534  & \textbf{0.616}  & 15.09 \\
                                     & NDCG@10 & 0.147 & 0.230 & 0.233 & 0.295 & 0.304  & 0.374  & 0.410  & \textbf{0.435}  & 4.88  \\ \hline
\end{tabular}
\end{table*}

The POP model, which is the most basic recommendation model, does not make use of historical interaction information between users and items, so it has the worst performance in all datasets. Compared with the POP model, the performance of the BPR is better, but it is not as good as the NeuMF since NeuMF uses neural networks to simulate the complex interaction relationships. From this set of comparative experiments, we can see that considering the interaction information can improve the recommendation performance, and the neural network can better simulate complex interactions.

Among the recommendation models which use neural network for reasoning, models using the sequence information between nodes in HIN, such as RNN4rec and CNN4rec, have better performance than NeuMF, which does not take sequences into consideration. Performance improvement is especially obvious on sparse data sets, which demonstrates that considering heterogeneous networks can alleviate the problem of data sparsity. Besides, the performance of CNN4rec is not as good as that of RNN4rec. This phenomenon may vary due to the limited number of path length in our sequence modeling. LSTM can solve the gradient vanishing problem in learning long-term dependencies. It can better learn the sequence information through the memory function. Compared with the ABLAH model's recommendation performance, RNN4rec is worse, indicating that the bidirectional LSTM can better learn the representation of the node itself. It demonstrates the effectiveness of the attention-based bidirectional LSTM and adversarial learning in the path inference process. 

As can be seen from the experimental results, we can conclude that ABLAH performs best among all models on three datasets. There is an increase of 17.28\% in HR@10 and 28.12\% in NDCG@10 (on average) against the strongest baselines.  

\subsection{Ablation Study (RQ2)}
To better understand the impact of each key component of ABLAH on the recommendation performance (attention mechanism (AM)) and reveal the vital role of adversarial learning (adversarial regularization term (ART)), we perform ablation experiments on three datasets. Table~\ref{tab:tab3} shows the results of ABLAH and its variants while keeping the hyper-parameters at their optimal setting. We introduce the variants of ABLAH and analyze their effects as follows:

\begin{enumerate}[\textbullet]
\item \textbf{w/o AM}: It is the variant model of ABLAH that lacks the attention mechanism to reason in HIN. We compare with it to validate the benefits of attention mechanism in the proposed model. The performances show that removing the attention mechanism causes a decline in ABLAH's performance on three datasets. Without the attention mechanism, the importance of each node is considered the same.

\item \textbf{ABLH}: It is the basic model of ABLAH that only uses attention-based bidirectional LSTM to reason, without considering the adversarial regularization term. We compare with it to answer the question that how does the adversarial training module affect model performance. The results show that the performance of the ABLAH model is not as good as ABLH when the path length selection is small. Nevertheless, when the path length selected from the HIN becomes longer (i.e., Xiami, and Yahoo), the performance of the ABLAH is gradually better than ABLH. To verify the effect of path length on the two models, we vary the path length of Xiami datasets, as shown in Table~\ref{tab:tab4}. We can see that the recommendation model achieves the best results when the number of nodes $L$ is 5. With the increase of $L$, the performance of the model gradually improves. The performance begins to decline when $L$ exceeds the optimal value, which shows that too many nodes in a path would introduce additional information and more noise, and therefore eventually affect the vector representation of the nodes. However, with the increase of $L$, our model's performance remains stable without any sharp decline, which shows that our model is robust. Overall, adding the adversarial regularization term can indeed learn more accurate node vector representations and improve the robustness of the embedding representations.

\item \textbf{w/o AM and ART}: It is the variant model of ABLAH that only uses bidirectional LSTM to reason in HIN, without considering the attention mechanism and adversarial regularization term. We compare with it to validate the benefits of attention mechanism in the proposed model. We observe that the performance of this model is worse than ABLH and w/o AM, which also verifies the positive effects of attention mechanism and adversarial regularization term on the node representation learning process.
\end{enumerate}

\begin{table}[pos=htp]
\caption{Ablation analysis (HR@10) on three datasets. Bold score indicates the best performances.}
\label{tab:tab3}
\begin{tabular}{llll}
\hline
\multicolumn{1}{c}{\multirow{2}{*}{\textbf{Models}}} & \multicolumn{3}{c}{\textbf{Datasets}}                 \\ \cline{2-4} 
\multicolumn{1}{c}{}                                 & \textbf{Nowplaying} & \textbf{Xiami} & \textbf{Yahoo} \\ \hline
w/o AM         & 5.116          & 3.393          & 10.754          \\
ABLH           & 5.086          & 3.311          & 10.460          \\
w/o AM and ART & 4.927          & 3.193          & 10.231          \\
ABLAH          & \textbf{5.143} & \textbf{3.478} & \textbf{10.875} \\ \hline
\end{tabular}
\end{table}

\begin{table}[pos=htp]
\footnotesize
\caption{Performance comparison with respect to different maximum path length $L$ on Xiami dataset. Bold score indicates the best performances.}
\label{tab:tab4}
\begin{tabular}{lllllll}
\hline
\textbf{Models} & \textbf{Metric(\%)} & \textbf{2} & \textbf{3} & \textbf{4} & \textbf{5} & \textbf{6} \\ \hline
\multirow{2}{*}{ABLH}  & HR@10   & 1.212 & 2.355 & 3.140 & \textbf{3.311} & 3.183 \\
                       & NDCG@10 & 0.047 & 0.091 & 0.123 & \textbf{0.128} & 0.125 \\ \hline
\multirow{2}{*}{ABLAH} & HR@10   & 2.177 & 2.591 & 3.115 & \textbf{3.478} & 3.229 \\
                       & NDCG@10 & 0.091 & 0.114 & 0.125 & \textbf{0.134} & 0.128 \\ \hline
\end{tabular}
\end{table}

\subsection{Parameter Sensitivity Analysis (RQ3)}
In this section, we study the effect of different settings of parameters on the recommendation performance. When exploring the effect of one hyper-parameter on the performance of the model, we fix other hyper-parameters to the same value.

\begin{figure*}[pos=htp]
\centerline{\includegraphics[scale=0.45]{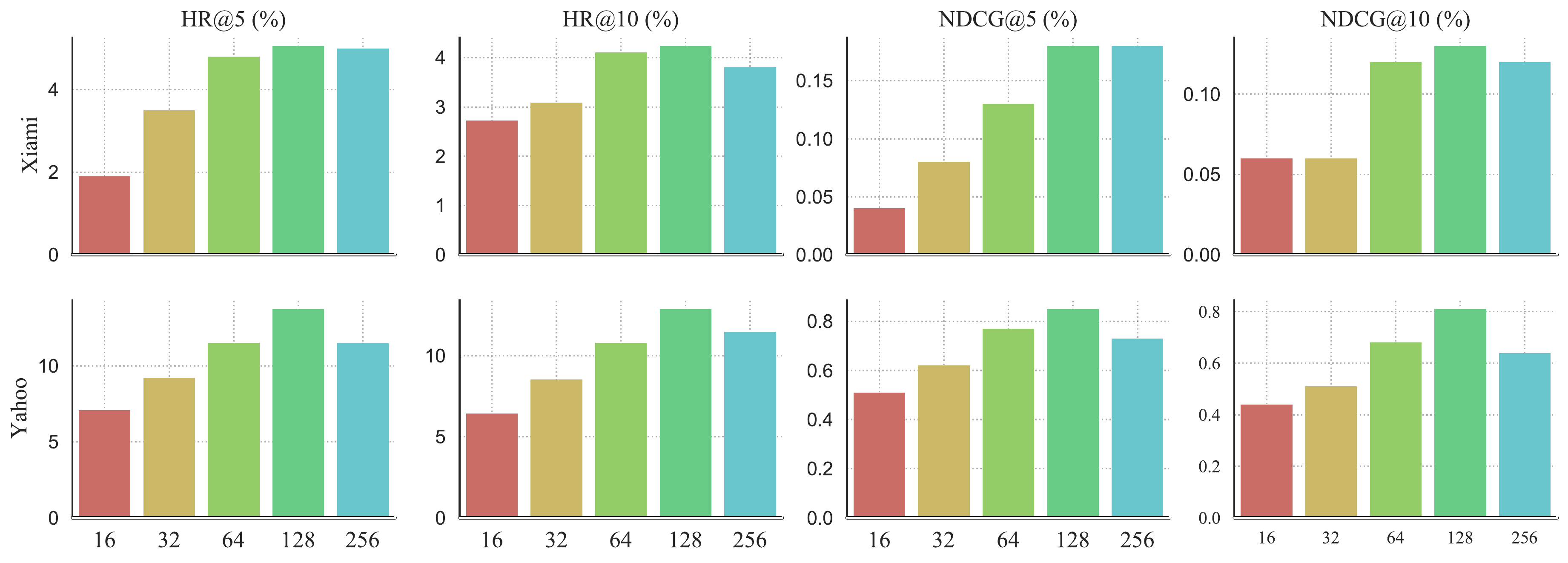}}
\caption{Effect of node embedding dimension $d$ on model.}
\label{f3}
\end{figure*}

\begin{figure*}
\centerline{\includegraphics[scale=0.45]{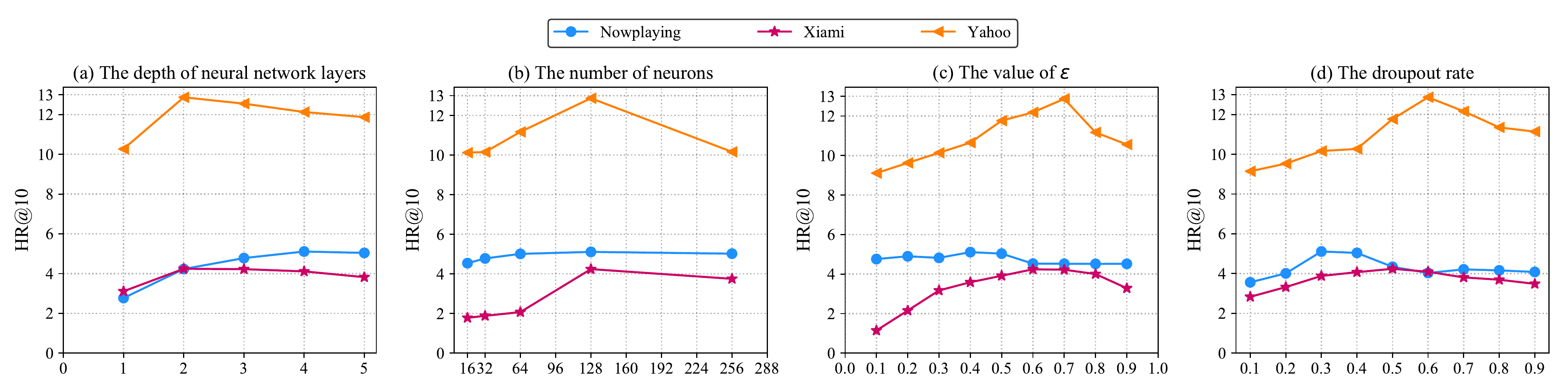}}
\caption{Effect of varying parameters on three datasets with HR@10.}
\label{f4}
\end{figure*}

\begin{enumerate}[\textbullet]
\item \textbf{The dimension of node embedding:} Fig.~\ref{f3} shows the change of model performance with respect to different dimension number from 16 to 256. It is obvious that model performance eventually converges with the rise of dimension. Larger embedding dimensions do not yield better model performance, especially in sparse data sets. Concretely, the optimal embedding dimensions is 128. Therefore, we set the dimension $d$ to 128 in other experiments.

\item \textbf{The depth of neural network layers:} In our model, we adopt neural networks to project the final state. Hence, the depth of the neural network is an important parameter. The optimal layer number of neural network is searched in {1, 2, 3, 4, 5} in our experiment. We present the results in Fig.~\ref{f4}(a). Through the experimental results, we can find that when the layer of the neural network is 4, the model achieves better performance in the Nowplaying dataset. Using 2 layers of neural networks yields the best performance in Xiami and Yahoo datasets, indicating that the sparser the data, the deeper the neural network layers should be.

\item \textbf{The number of neurons:} Besides, we analyze the number of neurons in each layer of the neural network, and the number of neurons in each layer is tuned among {16, 32, 64, 128, 256}. As shown in Fig.~\ref{f4}(b), when the number of neurons is 64, our model achieves the best performance. The HR@10 increases first and then decreases with the number of neurons, so we choose 128 as the number of neurons for the Yahoo and Xiami dataset.

\item \textbf{The value of $\epsilon$:} Next, we set the dimension of node embedding as 128 and tune the value of $\epsilon$, which is used to control the regularization term. We examine how it influences the model performance, when it varies from 0.1 to 1.0. As shown in Fig.~\ref{f4}(c), the optimal performance is obtained when $\epsilon$ is near 0.4 in the Nowplaying dataset. When the value of $\epsilon$ is 0.6, our model achieves the best performance in the Xiami and Yahoo dataset.

\item \textbf{The dropout rate:} Finally, we study the influence of dropout rate on model performance. Fig.~\ref{f4}(d) shows the results with respect to different dropout rate from 0.1 to 0.9 on three datasets. The Nowplaying dataset prefers a small value of 0.3. For the more sparse dataset (e.g., Xiami and Yahoo), the best performances are achieved when dropout rate is set to larger value, concretely 0.5 and 0.6  respectively. We can also see that the recommendation performance gradually decreases when the dropout rate becomes larger. We reckon that the model may suffer from under-fitting when the drop rate is too large.
\end{enumerate}

\subsection{Case Studies (RQ4)}
\begin{figure}[pos=htp]
\centerline{\includegraphics[scale=1.1]{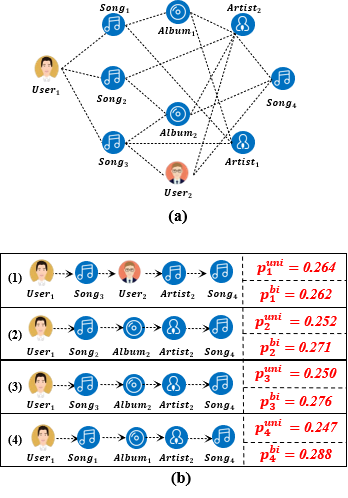}}
\caption{Visualization of four paths with prediction scores for the user of $User_{1}$ in Xiami dataset. The prediction scores are normalized for illustration.}
\label{f5}
\end{figure}

It is innovative that our model uses attention-based bidirectional LSTM to infer the node sequences over HIN. To demonstrate the feasibility of our model on explainable recommendation task, we randomly select a user $User_{1}$ from the Xiami dataset and show its 4 paths to song $Song_{4}$. As shown in Fig.~\ref{f5} (a), we can see that $Song_{4}$ is related to $Album_{2}$, $Artist_{2}$, and $User_{2}$. Through the display of different paths, we find that the paths describes the connectivity between $User_{1}$ and $Song_{4}$ from different perspectives, which can be seen as the reason why songs are recommended to users. Based on this, the target song is more persuasive when it is recommended to the user.

As shown in Fig.~\ref{f5}(b), we select 4 paths from the HIN about Xiami and calculate the weights of different paths from two ways: bidirectional and unidirectional. The weight of the path calculated by bidirectional is generally higher than that calculated by unidirectional, and the path (4) has the highest probability. Therefore, when the model recommends the $Song_{4}$ to the $User_{1}$, the reason for the recommendation can be shown to the user at the same time. For example, the user, who have listened $Song_{1}$, may be also interested in the $Song_{4}$, which belongs to the same album and same singer as $Song_{1}$. However, if we use unidirectional method, the model will recommend $Song_{4}$ to $User_{1}$ according to path (1), which shows that considering the bidirectional relationships of the nodes could better capture the complex nodes sequences.

\section{Conclusion \& Future Work}
In this paper, we introduce an attention-based bidirectional LSTM with adversarial learning for path-based reasoning over HIN, named ABLAH, to capture the complex node sequence information and highlight the different importance of nodes in HIN. As for noisy relationships in the node sequence, we use the adversarial regularization term to learn more accurate node vector representations and improve the robustness of the embedding representations. Extensive experimental results on three real-world datasets show that the superiority of our model compared to other state-of-art baselines. The case study validates the feasibility of the proposed model on explainable recommendation task.

There are still several directions to be explored in the future. A valuable one is to consider the position information of nodes in the sequence, instead of merely simulating the sequence of paths through the model. Another interesting direction is to consider the multiple types of edges and different attributes of nodes in HIN.

\section{Acknowledgements}
This research is supported by the National Key Research and Development Program of China (2018YFB1403600 and 2018YFF0214706), the Graduate Scientific Research and Innovation Foundation of Chongqing, China (CYS19028), and the Fundamental Research Funds for the Central Universities of Chongqing University (2019CDCGJG331).

\bibliographystyle{cas-model2-names}

\bibliography{cas-refs}

\printcredits
\vskip3pt

\end{document}